\documentclass[sigconf,screen, hidelinks]{acmart}

\AtBeginDocument{%
  }

\setcopyright{acmlicensed}
\copyrightyear{2025}
\acmYear{2025}
\acmDOI{XXXXXXX.XXXXXXX}
\acmConference[MM '25]{Proceedings of the 33rd ACM International Conference on Multimedia}{October 27--31,
  2025}{Dublin, Ireland}
\acmISBN{XXX-X-XXXX-XXXX-X/XXXX/XX}

\acmSubmissionID{6278}

\usepackage{epsfig}

\usepackage{amsmath}
\usepackage{amssymb}
\usepackage{color}
\usepackage{url}
\usepackage{xspace}
\usepackage{multirow}

\usepackage{graphicx} 
\usepackage{subcaption}
\usepackage{url}
\usepackage{tabularx,booktabs}
\newcolumntype{Y}{>{\centering\arraybackslash}X}
\newcolumntype{R}{>{\raggedleft\arraybackslash}X}
\usepackage{arydshln}
\usepackage{tabu}
\usepackage{ctable}
\usepackage{enumerate,multicol}
\usepackage{multirow}
\usepackage{multicol}
\usepackage{paralist}
\usepackage{array}
\usepackage{gensymb}
\usepackage{float}
\usepackage{caption}
\usepackage{balance}
\usepackage{xspace}
\usepackage{enumerate,multicol}
\usepackage{algorithmic}
\usepackage{amsmath} 
\usepackage[linesnumbered,ruled,vlined]{algorithm2e}
\usepackage{siunitx}

\usepackage[nolist]{acronym}
\usepackage[inline]{enumitem}
\usepackage{siunitx}
\begin{acronym}
    \acro{QoE}{Quality of Experience}
    \acro{AR}{Augmented Reality}
    \acro{VR}{Virtual Reality}
    \acro{AVP}{Apple Vision Pro}
    \acro{HMD}{Head-Mounted Display}
\end{acronym}

\begin{document}

\title{SVD: Spatial Video Dataset}

\author{M.H. Izadimehr}
\affiliation{%
  \institution{University of Klagenfurt}
  \city{Klagenfurt}
  \country{Austria}}

\author{Milad Ghanbari}
\affiliation{%
  \institution{University of Klagenfurt}
  \city{Klagenfurt}
  \country{Austria}}

\author{Guodong Chen}
\affiliation{%
  \institution{Northeastern University}
  \city{Boston}
  \country{USA}}

\author{Wei Zhou}
\affiliation{%
  \institution{Cardiff University}
  \city{Cardiff}
  \country{UK}}

\author{Xiaoshuai Hao}
\affiliation{%
  \institution{Beijing Academy of Artificial Intelligence}
  \city{Beijing}
  \country{China}}

\author{Mallesham Dasari}
\affiliation{%
  \institution{Northeastern University}
  \city{Boston}
  \country{USA}}

\author{Christian Timmerer}
\affiliation{%
  \institution{University of Klagenfurt}
  \city{Klagenfurt}
  \country{Austria}}

\author{Hadi Amirpour}
\affiliation{%
  \institution{University of Klagenfurt}
  \city{Klagenfurt}
  \country{Austria}}  


\renewcommand{\shortauthors}{Izadimehr et al.}

\begin{abstract} 
Stereoscopic video has long been the subject of research due to its capacity to deliver immersive three-dimensional content across a wide range of applications, from virtual and augmented reality to advanced human–computer interaction. The dual‑view format inherently provides binocular disparity cues that enhance depth perception and realism, making it indispensable for fields such as telepresence, 3D mapping, and robotic vision. Until recently, however, end‑to‑end pipelines for capturing, encoding, and viewing high‑quality 3D video were neither widely accessible nor optimized for consumer‑grade devices. Today’s smartphones—such as the iPhone Pro—and modern \acp{HMD}—like the \ac{AVP}—offer built‑in support for stereoscopic video capture, hardware‑accelerated encoding (e.g., HEVC/x265), and seamless playback on devices like the Apple Vision Pro and Meta Quest 3, requiring minimal user intervention. Apple refers to this streamlined workflow as spatial Video. Making the full stereoscopic video process available to everyone has made new applications possible. Despite these advances, there remains a notable absence of publicly available datasets that include the complete spatial video pipeline on consumer platforms, hindering reproducibility and comparative evaluation of emerging algorithms.

In this paper, we introduce SVD, a spatial video dataset comprising 300 five-second video sequences— 150 captured using an iPhone Pro and 150 with an \ac{AVP}. Additionally, 10 longer videos with a minimum duration of 2 minutes have been recorded.
The SVD dataset is publicly released under an open‑access license to facilitate research in codec performance evaluation, subjective and objective quality of experience (QoE) assessment, depth‑based computer vision, stereoscopic video streaming, and other emerging 3D applications such as neural rendering and volumetric capture. Link to the dataset: \url{https://cd-athena.github.io/SVD/}

\end{abstract}




\maketitle

\section{Introduction}
\label{sec:1}

Immersive media technologies~\cite{van_der_hooft_tutorial_2023} are redefining how digital content is experienced by delivering more realistic and visually compelling representations of scenes. Advances in virtual~\cite{wohlgenannt_virtual_2020,anthes_state_2016}, augmented~\cite{carmigniani_augmented_2011}, and mixed reality~\cite{speicher_what_2019} have driven the development of high-resolution \ac{HMD}~\cite{dunphy_integrating_2024,cheng_first_2024}, spatial audio integration, and improved stereoscopic rendering. These technologies are enabling deeply engaging experiences across domains such as entertainment, education, and visual communication, where realism and a strong sense of presence are essential.

\begin{table*}[h!]
    \centering
    \centering
    \caption{Overview of stereoscopic video datasets.}
    \begin{tabular}{lcccc}
        \toprule
        \textbf{Dataset Name} & \textbf{Year} &  \textbf{Resolution} & \textbf{Description} \\ 
        \midrule
        KITTI Stereo 2012~\cite{geiger_are_2012} & 2012  & 1226$\times$370 & Outdoor driving scenes \\ 
        KITTI Stereo 2015~\cite{menze_object_2015} & 2015  & 1242$\times$375 & Outdoor driving scenes with dynamic scenes with objects \\ 
        SceneFlow~\cite{mayer_large_2016} & 2016 &  960$\times$540 & Synthetic stereo sequences \\ 
        MPI-Sintel~\cite{butler_naturalistic_2012} & 2012 & 1024$\times$436 (24fps) & Synthetic scenes with complex motion and visual effects \\
        RMIT3DV HD 3D Video~\cite{cheng_rmit3dv_2012} & 2012 & 1920$\times$1080 (25fps)& Thirty-one diverse urban scenes \\ 
        EPFL MMSPG 3DVQA~\cite{goldmann_comprehensive_2010} & 2010  & 1920$\times$1080 (25fp) & Six high-quality visual variations \\ 
        Stereo Video Database~\cite{corrigan_video_2010} & 2010 &  1920$\times$1080 (25fps) & Stereo cinema post-production \\ 
        NAMAD3D~\cite{urvoy_nama3ds1-cospad1_2012} & 2012 &  1920$\times$1080(25fps) & Natural 3D scenes with twin-lens camera \\ 
        SVSR-Set~\cite{imani_new_2022} & 2022 & 1920$\times$1080 (30fps) & Indoor/outdoor with varied motion and lighting \\
        \midrule
        \multirow{2}{*}{SVD (Ours)} & 2025 & 1920$\times$1080 (30fps) & Indoor and outdoor, captured with iPhone Pro \\
                                   & 2025 & 2200$\times$2200 (30fps) & Indoor and outdoor, captured with \ac{AVP} \\
        \bottomrule
        
 \end{tabular}
    
    \label{tab:datasets}
\end{table*}

A key component of immersive media is stereoscopic video, which enhances realism by replicating the way human vision perceives depth through binocular disparity. In practice, this involves capturing two slightly offset views of a scene—one corresponding to the left eye and one to the right—using a dual‐lens or two‐camera rig that is carefully calibrated to maintain known baseline distance and optical parameters. During capture, precise synchronization and geometric calibration ensure that corresponding pixels in each view lie on the same epipolar line, facilitating accurate disparity estimation. At playback, specialized display technologies present each view to the appropriate eye. The human visual system then fuses these two images, leveraging small interocular differences to reconstruct a coherent depth map and evoke a convincing sense of three‐dimensional space. 

Despite its clear benefits for depth perception, stereoscopic video production has historically been constrained by increased capture complexity, the need for rigorous calibration, higher data rates to accommodate dual streams, and display hardware requirements that have, until recently, limited its adoption in consumer and broadcast contexts. 

Recently, this barrier has been significantly lowered through consumer devices that support native stereoscopic video workflows. Smartphones such as the iPhone Pro now offer built-in dual-camera setups for spatial video capture, while headsets like the \ac{AVP} and Meta Quest 3 provide native playback support. These devices also include hardware-accelerated encoding, enabling efficient compression using modern codecs like HEVC (x265). Apple has introduced the term \textbf{spatial} video to describe this tightly integrated pipeline from capture to playback, which allows users to create and experience 3D content with minimal technical effort.

While there are many well-established 2D video datasets~\cite{abu-el-haija_youtube-8m_2016, wang_youtube_2019, amirpour_vcd_2022}, the availability of high-quality stereoscopic video datasets has remained limited. This scarcity is largely due to the challenges associated with stereo video capture, the lack of accessible stereoscopic displays, and the need for optimized stereo video encoders. However, with recent advancements in capture technologies and wider availability of immersive displays, these barriers have significantly diminished. To drive research in stereoscopic video processing, we introduce the Spatial Video Dataset (SVD)—a comprehensive collection of high-quality stereoscopic video clips captured using the latest iPhone 16 Pro and \ac{AVP} devices. The dataset comprises 150 short 5-second videos from each device, along with 10 long-form sequences captured with both, covering a diverse range of indoor and outdoor environments, varied motion dynamics, and unique capture scenarios. SVD is specifically designed to support a broad spectrum of applications, including stereoscopic image and video coding, streaming, Quality of Experience (QoE) assessment, and stereoscopic image and video quality evaluation, providing researchers with a powerful resource for advancing immersive media technologies.

\section{Related work}
In this section, we introduce stereoscopic video datasets from the literature. The KITTI Stereo 2012 dataset~\cite{geiger_are_2012} serves as a key benchmark for stereo vision in autonomous driving. It contains stereo videos of road scenes captured from a calibrated pair of cameras mounted on a car. It includes 194 training and 195 test scenes with resolutions of 1226$\times$370, captured in outdoor environments with high-resolution stereo cameras. 

The KITTI Stereo 2015 dataset~\cite{menze_object_2015} builds upon its predecessor by adding 200 training and 200 test scenes with a resolution of 1242$\times$375 in dynamic environments with moving objects, enhancing its relevance for real-world driving scenarios.

SceneFlow~\cite{mayer_large_2016} provides a dataset containing synthetic stereo videos with a resolution of 960$\times$540. The RMIT3DV HD 3D Video database~\cite{cheng_rmit3dv_2012} is a comprehensive dataset designed to represent diverse content and visual conditions for various research applications. It comprises 31 stereoscopic video sequences filmed across multiple locations at RMIT University and Melbourne CBD, with durations ranging from 17 seconds to 2.5 minutes. All videos are recorded using a stereoscopic camera (Panasonic AG-3DA1) in 1920×1080  resolution with 10-bit YUV 4:2:2 encoding at 25fps, ensuring high visual fidelity and uncompressed quality. This dataset is particularly valuable for studies involving stereo video quality assessment, disparity estimation, and 3D visual analysis, providing high-resolution, uncompressed stereoscopic content for reliable experimental evaluation. 

The MPI-Sintel dataset~\cite{butler_naturalistic_2012}, derived from the open-source animated film Sintel, is a widely used benchmark originally developed for optical flow evaluation but also highly relevant for stereoscopic research. It includes stereo video pairs rendered at a resolution of 1024$\times$436 with rich visual effects such as motion blur, specular reflections, and atmospheric conditions, closely mimicking real-world scenes. Despite being synthetic, its image and motion statistics align well with those of natural videos, making it a credible proxy for stereo vision tasks. With dense ground truth, multiple rendering passes, and long sequences, MPI-Sintel provides a flexible and reproducible resource for benchmarking stereo matching, disparity estimation, and depth-aware video analysis.

The EPFL MMSPG HD 3D Video Database (3DVQA)~\cite{goldmann_comprehensive_2010} is a well-known dataset comprising six stereoscopic video scenes, each lasting 10 seconds and capturing a variety of colors, textures, motion, and depth variations. Recorded in 1920$\times$1080 resolution at 25fps, the videos are stored in AVCHD format and compressed with MPEG-4 AVC/H.264 at 24Mbps. Despite its compression, 3DVQA preserves high visual quality, making it an ideal resource for stereo video quality assessment, disparity estimation, and depth-aware encoding techniques. Its controlled yet diverse visual content supports reliable benchmarking in 3D video quality evaluation and computer vision research. 

The Stereo Video Database~\cite{corrigan_video_2010} is specifically designed as a test resource for research and development in stereo cinema post-production. It features a diverse collection of sequences shot in both indoor and outdoor environments under controlled and uncontrolled lighting conditions, capturing various real-world scenarios. The footage includes both steadicam and tripod-based shots, providing different levels of motion dynamics. The experimental setup employs a dual-camera rig with two Iconix HD-RH1 cameras mounted on an Inition `bolt' side-by-side rig. Data is recorded using Flash XDR units in 4:2:2 XDCAM format with the xd5e codec at a bitrate of 100 Mbps. All sequences are captured in 1920$\times$1080 resolution at 25fps, ensuring high-quality stereoscopic content suitable for post-production analysis and stereo depth processing. 

\begin{figure*}[t]
    \centering
    \includegraphics[width=1\textwidth]{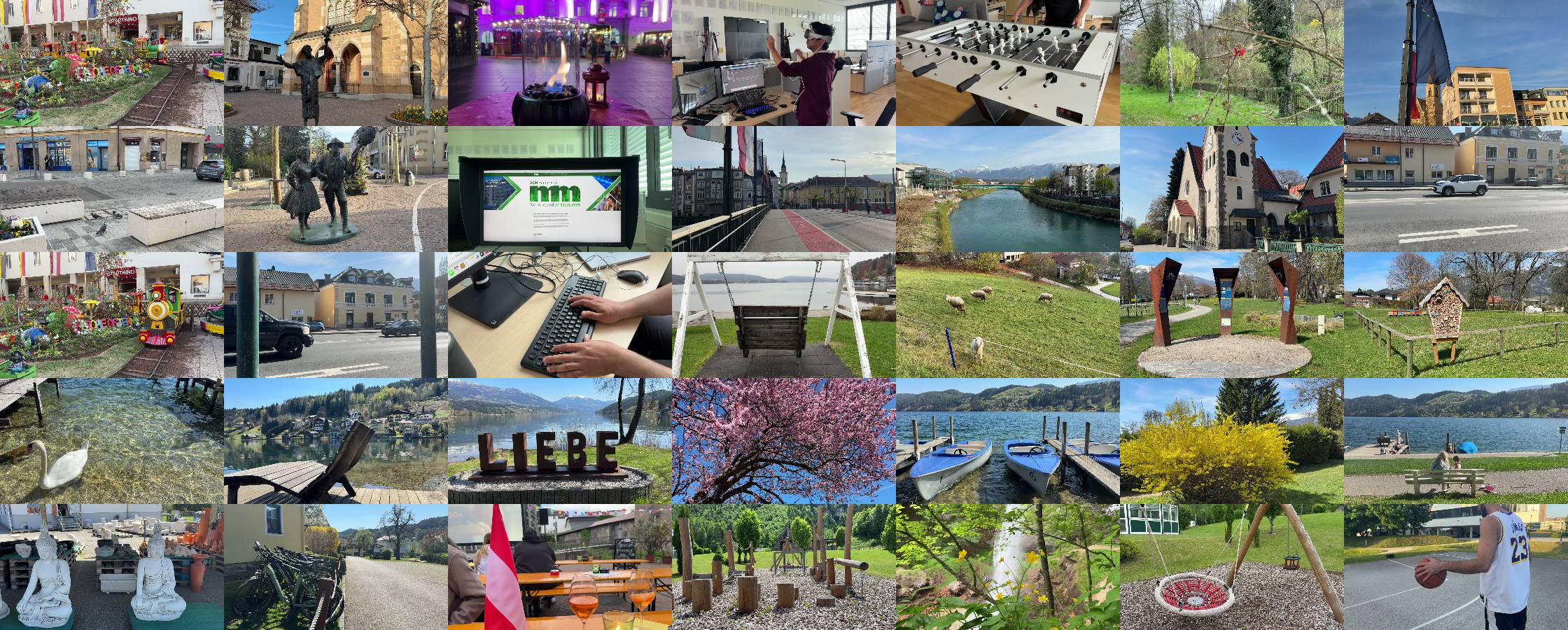} 
    \caption{Grid of the first frames from 35 randomly selected spatial videos recorded using the iPhone Pro. }
    \label{fig:grid}
\end{figure*}

In NAMAD3D~\cite{urvoy_nama3ds1-cospad1_2012}, the sequences were captured using a Panasonic AG-3DAIE twin-lens camera, which features two synchronized lenses with a 60 mm separation, closely matching the human interpupillary distance for natural-looking 3D content. The sequences are recorded in 1920×1080 at 25fps. When feasible, uncompressed dual SDI streams were sent to a Clearview Extreme system for high-quality recording, applied to sequences like Barrier gate, Hall, News report, Phone call, Soccer, Tree branches, and Umbrella. In cases where streaming to Clearview was impractical, like for Basket, Boxer, and Lab, the content was saved directly onto SD cards in H.264/AVC High-Profile format at a maximum bitrate of 24 Mbps (average 21 Mbps). 

The SVSR-Set~\cite{imani_new_2022} dataset consists of 71 stereo videos captured with a ZED 2 stereoscopic camera. Videos are recorded in 1920$\times$1080 resolution at 30fps for a duration of 20 seconds and are available in .svo and .avi formats. The dataset includes a wide range of indoor and outdoor settings, with variations in motion levels and illumination conditions. To ensure accuracy, the camera underwent a detailed calibration process to correct potential shifts in its internal parts. The calibration file, generated once and reused for all recordings, contains the exact locations of the left and right cameras and their optical properties.

\section{Spatial Video Dataset}

In this section, we introduce our dataset, named SVD (Spatial Video Dataset), which contains 310 stereoscopic video sequences captured using both the iPhone Pro and the Apple Vision Pro (\ac{AVP}). We recorded a diverse subset of spatial video sequences with each device, covering a variety of indoor and outdoor scenarios to ensure content variability across lighting conditions, environments, and motion characteristics. Specifically, we captured 150 short video clips of 5 seconds each, along with 10 longer sequences per device, tailored to streaming-oriented use cases. A grid of the first frames from 35 randomly selected videos recorded with the iPhone Pro setup is shown in Fig.~\ref{fig:grid}, providing a visual summary of the diversity within our dataset.

We begin by detailing the camera configurations and recording capabilities of the two devices, highlighting their roles in enabling high-quality spatial video capture without the need for external calibration or rigs. We then describe the set of low-level features extracted from the dataset, including spatial and temporal complexity, colorfulness, and luminance statistics, providing a quantitative characterization of the visual content. 

\subsection{Camera Configurations}
Stereoscopic video capture has traditionally required complex and carefully calibrated hardware configurations~\cite{lee_novel_2000,tzavidas_multicamera_2005}. Conventional stereoscopic rigs often employ two physically separate cameras mounted on a rail or a rigid rig. These setups typically required manual alignment, synchronization, and post-processing to ensure temporal and geometric consistency between the two video streams. Moreover, ensuring perfect lens matching, exposure control, and white balance between the cameras was necessary to avoid visual discomfort or depth perception errors during playback.

While effective in controlled studio environments, these traditional systems were bulky, expensive, and impractical for casual or mobile content capture. Their complexity created a barrier to the broader adoption of stereoscopic video, particularly among non-professional users.

Recent innovations in consumer electronics have dramatically simplified stereoscopic video capture. Modern devices such as the iPhone  Pro and the \ac{AVP} integrate dual-camera systems and advanced computational photography pipelines that enable spatial video recording without the need for external rigs or manual calibration.

\subsubsection{iPhone  Pro Camera System}
The iPhone  Pro features native spatial video recording by utilizing its precisely calibrated wide and ultrawide rear cameras, which are spaced 19.2 mm apart to produce depth cues suitable for small screens and head-mounted displays (HMDs). Apple’s spatial video system integrates real-time depth estimation, optical stabilization, and synchronized exposure control to ensure high-quality stereo capture. Videos are recorded in 1080p at 30 fps in standard dynamic range (SDR) and encoded in HEVC with stereoscopic metadata, enabling seamless playback on devices like the \ac{AVP}.  In spatial video capture on the iPhone Pro, the concept of the "hero eye" refers to the primary camera—the Wide (1x) lens—that records the main view. This lens provides the higher-quality image, while the Ultra Wide (0.5x) lens captures a secondary view that is cropped and scaled to match the primary perspective.

\subsubsection{Apple Vision Pro Camera System}
The \ac{AVP} represents a significant advancement in immersive media, offering both playback and recording capabilities for spatial video. Equipped with a stereoscopic 3D main camera system featuring 18 mm lenses with an ƒ/2.00 aperture, the AVP captures spatial videos at a resolution of 2200 × 2200 pixels per eye at 30 frames per second in SDR. 
Spatial videos on the AVP are encoded using the Multiview High-Efficiency Video Coding (MV-HEVC) format. This format stores stereoscopic views in separate layers—one for each eye—within a single video file, accompanied by spatial metadata that enables immersive playback experiences. Table~\ref{tab:spatial-video-comparison} compares the spatial video recording capabilities of the iPhone  Pro and the \ac{AVP}.

\begin{table*}[ht]
  \centering
  \caption{Spatial Video Recording: iPhone 16 Pro vs.\ \ac{AVP}}
  \label{tab:spatial-video-comparison}
  \resizebox{0.9\linewidth}{!}{%
  \begin{tabular}{@{} l l l @{}}
    \toprule
    \textbf{Feature}                         & \textbf{iPhone 16 Pro}                                          & \textbf{Apple Vision Pro}                                      \\
    \midrule
    Resolution \& Frame Rate                 & 1920$\times$1080\,px @ 30fps (SDR)                            & 2200$\times$2200\,px @ 30fps (SDR)                            \\
    Video Format                 & MV-HEVC                                      & MV-HEVC                                       \\
    
    Horizontal Field of View (FOV)           & 63.4°                                                            & 71.6°                                                          \\
    Baseline (Interaxial Distance)           & 19.2mm                                                        & 63.8mm                                                        \\
    Hero Eye Concept      & Yes (“hero eye” from Wide camera); 
    & No (both eye streams equal quality)                            \\
    Recording Orientation Requirement      & Landscape                                                       & -                                                 \\
    \bottomrule
  \end{tabular}}
\end{table*}

\subsection{Low-Level Video Features}
For each video, we extract a comprehensive set of low‐level features on a per‐frame basis and include them alongside the original video in our released dataset. These features are widely used in video analysis and objective quality assessment, and they cover spatial, temporal, stereo-view, and perceptual dimensions. 

\subsubsection{Spatial Complexity}
Spatial complexity is a fundamental aspect of video content that significantly impacts both perceptual quality and compression efficiency. Scenes with high spatial detail—such as textures, edges, and fine patterns—are more challenging to compress without introducing visible artifacts, while simpler, smoother areas are easier to encode efficiently. For this reason, spatial complexity is widely used in video quality assessment and adaptive encoding strategies.

In our dataset, we quantify spatial complexity using two complementary features: Spatial Information (SI) and Spatial Complexity (SC). SI measures edge strength by applying a Sobel filter to each frame, capturing local contrast and sharpness. It is a well-established feature in standards such as ITU-T P.910 and correlates strongly with perceived detail. SC, on the other hand, is derived from the EVCA framework~\cite{amirpour_evca_2024} and operates in the DCT domain, capturing frequency-based spatial variation and block-level detail. It computes the spatial complexity by applying a weighted sum to the DCT coefficients of each block, where higher-frequency components are given greater emphasis to reflect the contribution of fine textures and detailed patterns within the frame.

\begin{figure*}[ht]
    \centering

    \begin{subfigure}[b]{0.48\linewidth}
        \includegraphics[width=\linewidth]{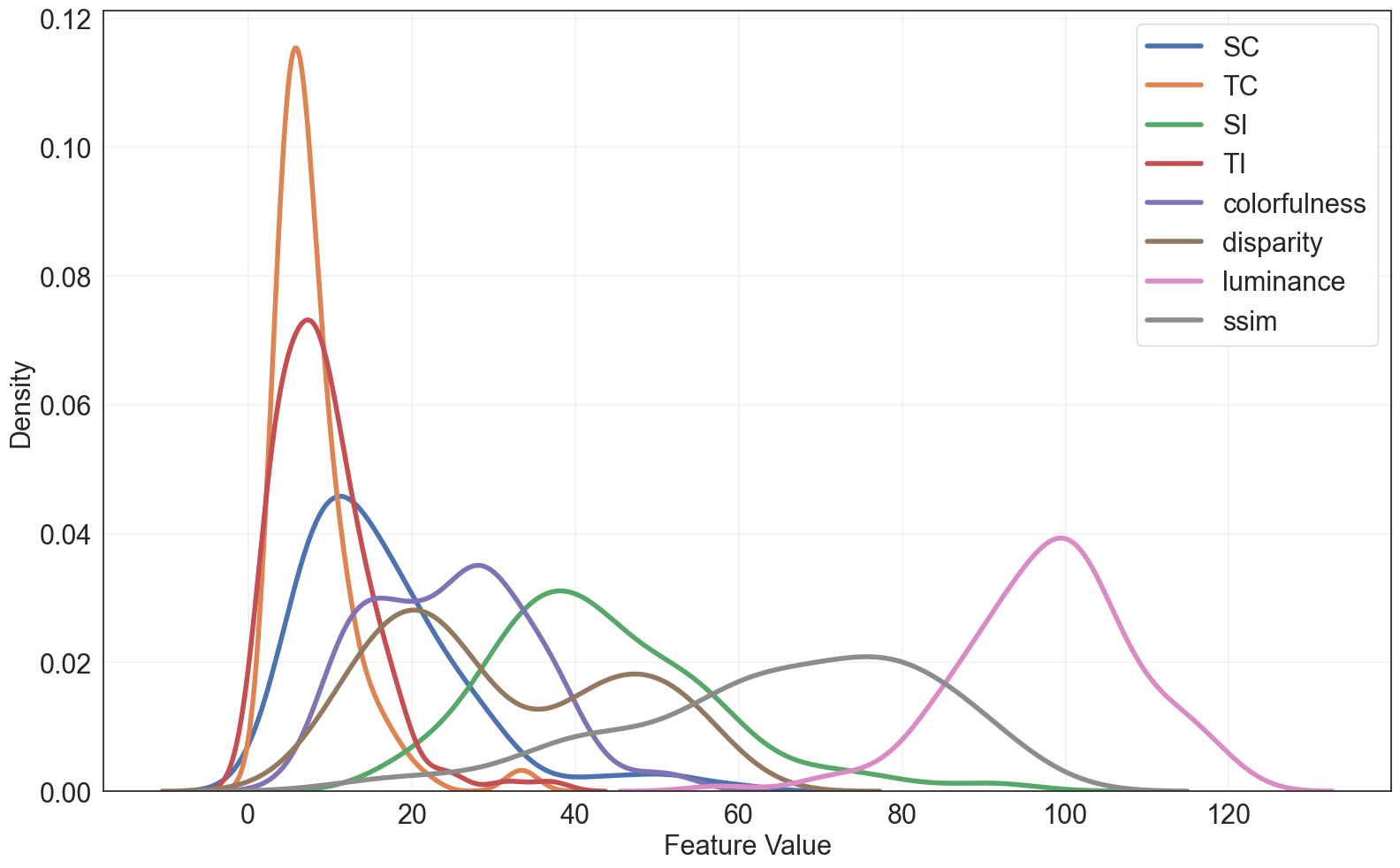}
        \caption{AVP Features}
        \label{fig:avp}
    \end{subfigure}
    \hfill
    \begin{subfigure}[b]{0.48\linewidth}
        \includegraphics[width=\linewidth]{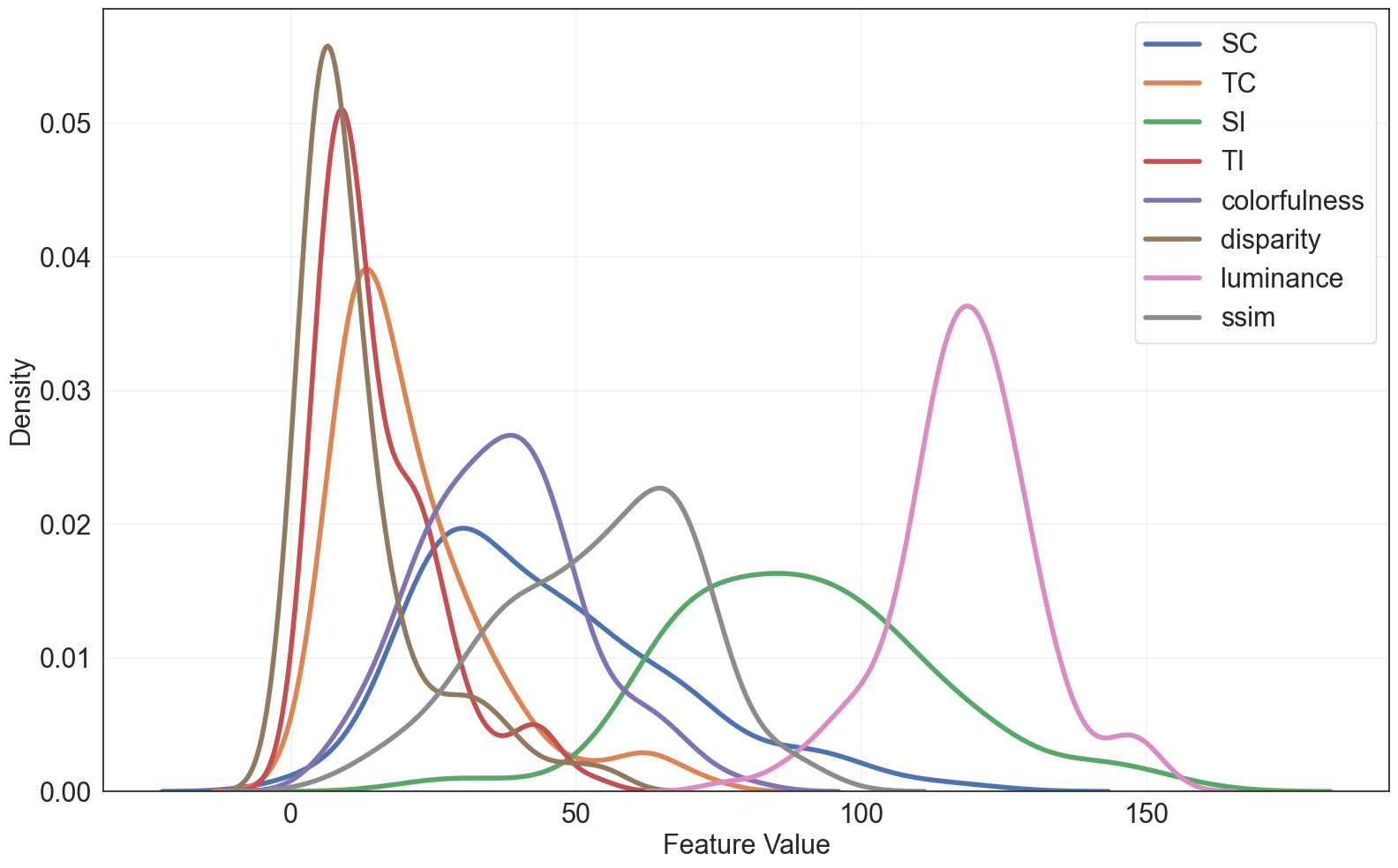}
        \caption{iPhone Features}
        \label{fig:iphone}
    \end{subfigure}

    \caption{Comparison of feature distributions from AVP and iPhone datasets.}
    \label{fig:feature_comparison}
\end{figure*}

\subsubsection{Temporal Complexity}
Temporal complexity reflects the amount of motion and dynamic change within a video, which significantly affects both perceived quality and compression performance. Videos with fast-moving objects, frequent cuts, or high activity between frames typically demand more resources for encoding and are more susceptible to motion-related artifacts.

To capture temporal complexity in our dataset, we use two complementary metrics: Temporal Information (TI) and Temporal Complexity (TC). TI is a well-established feature computed as the standard deviation of pixel-wise differences between consecutive frames, providing a frame-level measure of motion intensity. Higher TI values indicate stronger temporal variation, which is critical for tasks like motion-aware encoding, frame rate control, and adaptive streaming.

In addition to TI, we include TC, a motion-sensitive feature introduced in the EVCA framework~\cite{amirpour_evca_2024}. Unlike TI, which operates in the pixel domain, TC is calculated in the DCT domain by computing the sum of absolute differences (SAD) between the weighted DCT coefficients of corresponding blocks across consecutive frames. This weighting scheme emphasizes high-frequency components and thus captures subtle motion details and structural changes more effectively. TC has been shown to correlate more strongly with perceptual temporal complexity than earlier pixel-domain metrics. 

\subsubsection{Colorfulness}
Colorfulness is a perceptual attribute that reflects the intensity and diversity of colors within a video frame. It plays an important role in visual quality perception, content classification, and aesthetic evaluation. Videos with rich and varied colors tend to be perceived as more vivid and engaging, while those with dull or limited color ranges may appear flat or less appealing. In our dataset, we include a colorfulness metric introduced in~\cite{haskell_digital_2002}, which combines the mean and standard deviation of red-green and yellow-blue color differences. The colorfulness feature is computed for each frame of both left and right views, allowing for the analysis of color consistency across stereo pairs.

\subsubsection{Luminance}
In addition to spatial, temporal, and color features, we also include luminance-based metrics to capture the overall brightness and contrast characteristics of each frame. Specifically, we compute the mean and variance of the luminance (Y) channel for both left and right views. These features provide insight into lighting conditions, exposure balance, and perceptual contrast within the video, which can influence both encoding efficiency and visual quality perception.

\subsubsection{Disparity}
Disparity refers to the horizontal offset between corresponding points in the left and right views of a stereoscopic video, and it provides a key cue for depth perception. To capture disparity information in our dataset, we compute dense disparity maps for each video frame using the Semi-Global Block Matching (StereoSGBM) algorithm~\cite{hirschmuller_stereo_2008}, as implemented in OpenCV. This method balances local accuracy with global smoothness by aggregating matching costs along multiple paths, making it suitable for high-resolution stereo content.

\subsubsection{SSIM}

In addition to disparity, we compute the Structural Similarity Index (SSIM)~\cite{wang_image_2004} between the left and right views of each video frame to assess their perceptual correspondence. SSIM is a widely used image quality metric that evaluates luminance, contrast, and structural similarity, providing a more perceptually relevant comparison than pixel-wise differences. In our context, it serves as a complementary feature to disparity, offering a view-independent measure of stereo consistency. High SSIM values indicate strong structural alignment between the views, while lower values may signal mismatches, occlusions, or inconsistencies in stereo rendering.

Fig.~\ref{fig:feature_comparison} shows the distribution of the extracted low-level features for videos captured with the AVP and iPhone Pro devices using kernel density estimation (KDE) plots. All features—including spatial complexity, temporal complexity,  luminance, disparity, and SSIM (scaled by 100 for better visual representation)-are computed on a per-frame basis for both the left and right views. The values are then averaged across all frames to produce a single representative feature vector per video. These KDE plots highlight the differences in content characteristics and capture profiles between the two devices, offering insights into the diversity and quality of the dataset.

\begin{figure}
    \centering
    \includegraphics[width=1\linewidth]{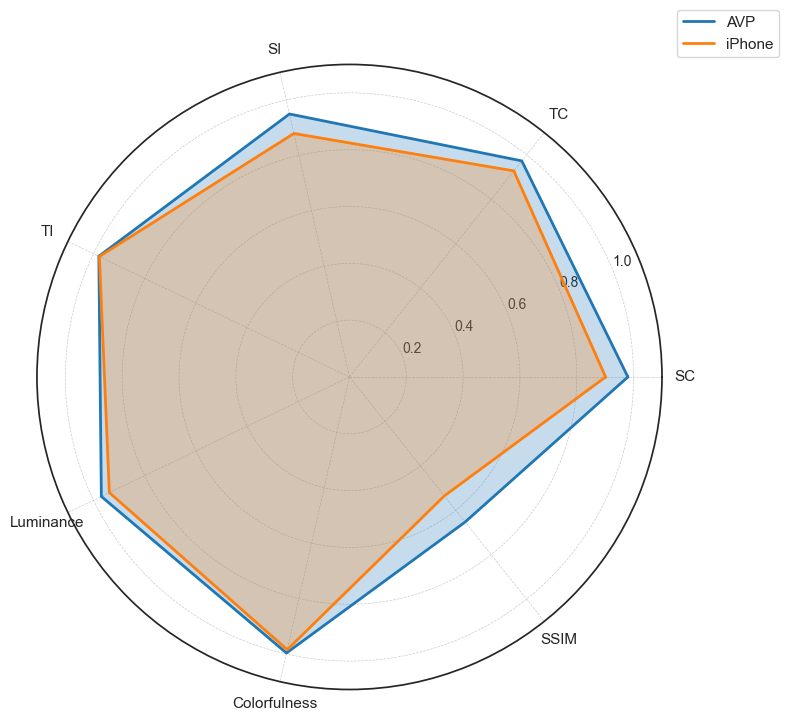}
    \caption{Correlation of low-level features and average SSIM between left and right views for spatial videos recorded with the iPhone Pro and \ac{AVP} (\ac{AVP}).}
    \label{fig:correlation}
\end{figure}

Fig.~\ref{fig:correlation} shows the \textbf{correlation} between corresponding low-level features extracted from the left and right views, along with the average SSIM values, for both the AVP and iPhone Pro devices. This analysis provides insight into the consistency of stereo content captured by each device. The results indicate that the AVP exhibits stronger correlations between views across most features, as well as higher SSIM scores, suggesting more consistent stereo alignment and better structural similarity. This highlights the superior stereo capture quality of the AVP compared to the iPhone Pro in our dataset.
The lower consistency observed in the iPhone Pro recordings may be attributed to the "hero eye" concept, where the Wide (1x) camera serves as the primary view and the Ultra Wide (0.5x) camera is cropped and aligned post-capture. This asymmetric processing can introduce disparities in quality and content between the two views.

\section{Potential Applications}
The rich set of features and high-quality stereoscopic content included in our dataset enables a wide range of research and development applications across multimedia, computer vision, and immersive media domains. Below, we outline several key areas where this dataset can be effectively leveraged.

\subsection{Codec Development and Comparison}

This dataset serves as a practical benchmark for codec development and evaluation, particularly for stereoscopic and multiview content. Earlier standards, such as MVC in H.264~\cite{vetro_overview_2011} and MV-HEVC in HEVC~\cite{tech_overview_2016}, introduced inter-view prediction to improve compression efficiency for stereo video. More recently, Apple adopted MV-HEVC for its Spatial Video format, and as of version 4.1, the x265 encoder added support for MV-HEVC, enabling optimized stereoscopic encoding within its efficient compression framework.

With rich diversity—including spatial and temporal complexity, disparity, and SSIM—our dataset allows for comprehensive codec comparisons in terms of rate-distortion performance, view consistency, and encoding speed. It also supports the evaluation of fast encoding algorithms and learning-based strategies for content-adaptive compression.

\subsection{Monoscopic-to-Stereoscopic Video}

Our dataset can be used to train and evaluate models that convert monoscopic (2D) videos into stereoscopic (3D) formats—an increasingly important task for supplying immersive content in AR/VR applications~\cite{zhang_spatialme_2024,yang_depth_2024}. As an inherently ill-posed problem, stereo conversion has evolved significantly with deep learning, progressing from early convolutional approaches to advanced diffusion-based models. These methods typically generate the right view from the left by estimating monocular depth and compensating for occluded regions through inpainting or generative synthesis. However, they often suffer from artifacts and lack control over structural accuracy. By offering high-quality stereo pairs, dense disparity maps, and perceptual similarity metrics such as SSIM, our dataset provides strong supervision and validation tools for improving the realism, consistency, and fidelity of stereoscopic view synthesis.

\subsection{Video Quality Assessment}
Our dataset is well-suited for conducting subjective quality assessments of stereoscopic video, thanks to its diversity in various features. This variability enables controlled experiments that evaluate how different content characteristics influence human perception of 3D video quality under various viewing conditions, including head-mounted displays and stereoscopic monitors~\cite{zhou_perceptual_2025}.
The outcomes of such subjective studies can be used to develop and validate both full-reference and no-reference video quality metrics tailored for stereoscopic content~\cite{wei_zhou_3d-hevc_2016}.

\subsection{Video Streaming}
The longer video sequences in our dataset make it particularly suitable for streaming applications, enabling realistic evaluations of adaptive delivery strategies over time~\cite{chen_spatial_2025,timmerer_http_2025}. These clips support research in content-aware bitrate ladder construction~\cite{menon_opte_2022}, where spatial, temporal, and disparity features can inform optimal quality tiers for stereoscopic video. The dataset also facilitates per-title encoding~\cite{amirpour_pstr_2021,amirpour_deepstream_2022,telili_convex_2025}, allowing encoding parameters to be tailored to individual content characteristics for improved compression efficiency and visual quality. Furthermore, it enables studies on Quality of Experience (QoE) in 3D streaming, including the effects of bitrate fluctuations, depth artifacts, and inter-view inconsistencies. By combining objective features with potential subjective evaluations, the dataset offers a comprehensive foundation for developing and testing adaptive streaming algorithms for stereoscopic and immersive video services.

\section{Conclusion}

We presented SVD, a publicly available spatial video dataset designed to support a broad range of research in stereoscopic and immersive media technologies. Captured using consumer-grade devices—namely the iPhone Pro and \ac{AVP}—the dataset includes both short and long-form high-quality stereoscopic video sequences, covering a wide range of real-world scenes. Alongside the raw videos, we provide a rich set of low-level features including spatial and temporal complexity, luminance, colorfulness, disparity, and inter-view SSIM, enabling in-depth analysis across multiple application domains.

SVD is specifically tailored for tasks such as codec development and benchmarking, monoscopic-to-stereoscopic video synthesis, video quality assessment (both subjective and objective), and adaptive streaming. Its inclusion of diverse content types, extended sequence durations, and per-frame metrics makes it an ideal resource for training, evaluating, and comparing algorithms in both traditional and emerging 3D video processing tasks. 

\newpage
\balance
\bibliographystyle{ieeetr}
\bibliography{files/ref,files/references}

\end{document}